\documentclass[apj]{emulateapj}
\usepackage{natbib}
\usepackage{color}
\begin{document}
\title{The Impact of the Spectral Response of an Achromatic Half-Wave Plate on the Measurement of the Cosmic Microwave Background Polarization}

\author{C.Bao\altaffilmark{1}, B.Gold\altaffilmark{1}, C.Baccigalupi\altaffilmark{2},
        J.Didier\altaffilmark{3}, S.Hanany\altaffilmark{1}, A.Jaffe\altaffilmark{4}, 
        B.R.Johnson\altaffilmark{3}, S.Leach\altaffilmark{2}, T.Matsumura\altaffilmark{5},
        A.Miller\altaffilmark{3} and D.O'Dea\altaffilmark{4}}
\altaffiltext{1}{University of Minnesota School of Physics and Astronomy,
Minneapolis, MN 55455}
\altaffiltext{2}{SISSA, Astrophysics Sector, via Bonomea 265, Trieste 34136, Italy}
\altaffiltext{3}{Columbia University, New York, NY 10027}
\altaffiltext{4}{Imperial College, London, SW72AZ, England, United Kingdom}
\altaffiltext{5}{High Energy Accelerator Research Organization (KEK), 1-1 Oho, Tsukuba, Ibaraki 305-0801, Japan}

\begin{abstract}
We study the impact of the spectral dependence of the linear polarization rotation induced by an
achromatic half-wave plate on measurements of cosmic microwave background polarization in the presence
of astrophysical foregrounds.
We focus on the systematic effects induced on the measurement of inflationary gravitational waves by
uncertainties in the polarization and spectral index of Galactic dust. We find that for the
experimental configuration and noise levels of the balloon-borne EBEX experiment, 
which has three frequency bands centered at 150, 250, and 410 GHz, a crude dust subtraction 
process mitigates systematic effects to below detectable levels for 10\% polarized dust and 
tensor to scalar ratio of as low as $r=0.01$. We also study the impact of uncertainties in the 
spectral response of the instrument. With a top-hat model of the spectral response for each 
band, characterized by band-center and band-width, and with the same crude dust subtraction process, 
we find that these parameters need to be determined to within 1 and 0.8~GHz at 150 GHz; 9
and 2.0~GHz at 250 GHz; and 20 and 14~GHz at 410 GHz, respectively. The approach presented in 
this paper is applicable to other optical elements that exhibit polarization rotation as a function
of frequency.
\end{abstract}


\section{Introduction}

The Cosmic Microwave Background (CMB) polarization field can be decomposed into two orthogonal E 
and B modes. On large angular scales the B-mode signal encodes information about inflation, a 
period of rapid expansion in the early universe \citep{Kamionkowski97,seljak97}. The signal is 
characterized by the tensor-to-scalar ratio $r$ which quantifies the relative strength of inflationary 
gravitational waves (IGW) and density perturbations generated by inflation. The level of the IGW signal 
encodes information about the energy scale at which inflation occurred. 
The current upper limit is $r~<~0.2$ \citep{WMAP7yr_Komatsu}. A number of experimental efforts are 
ongoing to search for the signal at levels as low as $r \sim 0.01$ over the coming years. On small 
angular scales, the B-mode signal is dominated by the `lensing signal', which results from gravitational 
lensing of the CMB photons by the large scale structure of the universe. The lensing converts E-mode to 
B-mode polarization \citep{Zaldarriaga1998}.

Galactic foregrounds are expected to be a source of confusion for measurements of the B-mode signal. 
Above 70~GHz the polarized emission from Galactic dust is predicted to dominate over much of the sky 
and be comparable to the IGW signal with an $r$ value of 0.1 or less even for the cleanest 
regions of the sky \citep{Page2007_WMAP3yrpol, Gold2009, Fraisse2011}. Therefore many experimental 
efforts plan to employ multiple frequencies which will enable foreground identification and subtraction. 

Some CMB polarimeters use a half-wave plate (HWP) to modulate the observed linear polarization, 
such as EBEX \citep{BRKspie}, SPIDER \citep{SpiderSPIE} and POLARBEAR \citep{PolarbearSPIE}. 
When observing at multiple frequency bands simultaneously, an achromatic half-wave plate 
(AHWP) can be used. An AHWP is a stack of monochromatic HWPs with a particular set of orientation
angles relative to each other \citep{Pancharatnam1955}. While an AHWP has a higher modulation efficiency 
across a broad frequency
range compared to a single HWP, it rotates the polarization angle of the incident light by an 
amount that depends on 
frequency \citep{AHWP_original,Matsumura09}. The amount of rotation depends on the construction 
parameters of the AHWP and on the spectrum and polarized intensity of the constituent signals, which 
for this paper are CMB and Galactic dust. With knowledge of the spectrum and relative 
polarization intensity, the amount of rotation can be calculated and corrected. However, while the 
spectrum of the CMB component is well known, that of dust is not. The polarized intensities of dust and 
CMB are also not well known. These uncertainties may pose challenges in the extraction of the 
underlying IGW signal. Various authors studied the impact of HWP non-idealities on measurements 
of CMB polarization \citep{Brown2009, Brian2010}. However, this particular frequency dependent rotation 
effect has not been studied in the context of B-mode measurements. The goal of this paper is to 
quantitatively assess this effect. For concreteness we adopt the AHWP model, frequency bands and 
approximate noise information that are applicable to the E and B experiment (EBEX) \citep{BRKspie}, a 
balloon-borne CMB polarimeter targeting the IGW signal at the $r \sim 0.04$ level. 

In Section 2 we describe the basic components of the simulation. Section 3 focuses on quantifying
the effect of rotation due to the AHWP in the 150 GHz band. In Section 4 we use multiple
frequency information to account for rotation due to Galactic dust.
In Section 5, we study the additional effects of uncertainties in the spectral response of
the instrument, and in Section 6 we make concluding remarks.

\section{Description of the Simulation}

\begin{figure}
 \plotone{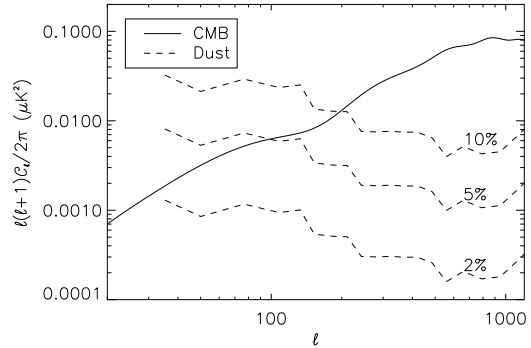}
 \caption{CMB (solid) and Galactic dust (dashed) B-mode power spectra at 150 GHz assuming an $r$ value of 
 	$0.05$. The spectra from dust are for the specific area of sky simulated in this work and are 
        given for three fractional polarization cases of 2, 5, and 10 \%. }
 \label{fig:cmb_dust_ps}
 \end{figure}

We simulate input Stokes $Q$ and $U$ signals due to the CMB and Galactic dust emission on a 
$10\degr \times 10\degr$ area of the sky centered on $( l, b ) = (252^{\circ} , -52^{\circ})$ in Galactic 
coordinates which is close to the center of the area targeted by EBEX. The maps are smoothed with an 
8\arcmin ~FWHM Gaussian beam then projected to a flat sky and pixelized with a square 6.9\arcmin~pixel.
Same simulations with a $20\degr \times 20\degr$ patch in the same region validate that 
conclusions presented in this paper do not depend on patch size. The input CMB polarization angular power 
spectra, including both the primordial and lensing signal, are generated with CAMB \citep{CAMB} using the 
best fit WMAP 7-year cosmological parameters \citep{WMAP7yr_Komatsu} and $r=0.05$, unless otherwise 
indicated. Our polarized foreground simulation follows the prescription detailed in \citet{Stivoli2010} 
and is briefly reviewed here. The dust intensity and its frequency scaling are given by `model 8' 
of~\citet{FDS99}. The dust polarization fraction is modeled for cases of 2\%, 5\%, 
and 10\%. A polarization fraction higher than 10\% would exceed the limit based on WMAP 
observations at intermediate and high Galactic latitudes \citep{Page2007_WMAP3yrpol, Gold2009}.
Both the dust polarization fraction and the frequency scaling are assumed to be uniform over the simulated
sky area. Observations suggest that this is a good approximation ~\citep{Planck2011,FDS99}. 
The pattern of the polarization angles on large angular scales ($l \lesssim 100$) is given by the
WMAP dust polarization template \citep{Page2007_WMAP3yrpol}. On smaller angular scales 
($l \gtrsim 100$) we add a Gaussian fluctuation power adopting a recipe first presented by 
\citet{Giardino2002_galpolmodel}.
Figure~\ref{fig:cmb_dust_ps} shows the power spectra of the CMB and 
of Galactic dust. For a level of 5\% fractional polarization the expected level of Galactic dust is 
comparable to the B-mode signal at $\ell=90$. 
  
To simulate the operation of the AHWP we use the Mueller matrix formalism as described by
~\citet{Matsumura09}. The level of input polarized signal
is calculated for each map pixel in 50 frequency bins for each of the experiment's three 
top-hat bands (see Table~\ref{tab:sim_pars}). For each map pixel the detected intensity
as a function of AHWP angle, which we call intensity vs. angle (IVA), is calculated for 
each frequency bin with an angular resolution 
of 0.05\degr\ and the total per-band IVA is the average of the 50 IVAs. The detected 
polarization angle, which is rotated relative to the input polarization angle, is encoded by the phase 
of the band-averaged IVA. To obtain the rotated map observed by the detector, we multiply each pixel of 
the input maps by a rotation matrix with the calculated phase of the band-averaged IVA. The frequency 
and IVA angular resolution are chosen to optimize computation time while giving negligible bias in the 
results. The construction parameters of the AHWP are given in Table~\ref{tab:sim_pars}.

\begin{table}[htbp]
 \begin{center}
 \begin{tabular}{|l|c|} 
 \hline
  Indices of refraction of AHWP      & $n_o = 3.047$, $n_e = 3.364$ \\\hline
  Thickness of each wave plate       & 1.69 mm  \\\hline
  150 GHz band                       & 133 - 173 GHz \\\hline
  250 GHz band                       & 217 - 288 GHz \\\hline
  410 GHz band                       & 366 - 450 GHz \\\hline
  Band shape                         & top-hat \\\hline
  Orientation angles of 5-stack AHWP & (0\degr, 25\degr, 88.5\degr, 25\degr, 0\degr) \\\hline
 \end{tabular}
\end{center}
\caption{AHWP and band parameters used in the simulations.}
\label{tab:sim_pars}
\end{table} 

\begin{figure}
 \plotone{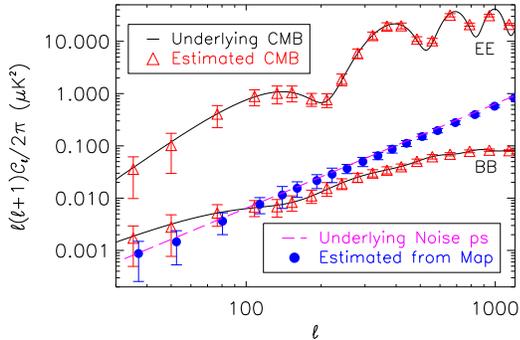}
 \caption{Validation of signal and noise power spectrum estimation. We generate 100 $Q$ and $U$ maps using
        an underlying power spectra (black solid curves) and use a flat-sky approximation to estimate 
        the power spectra (red triangles with error bars). The size of error bars agrees with predictions 
        for the contributions of cosmic and sample variance. We also make 100 noise only $Q$ and $U$ 
        realizations
        using white noise with RMS of 1 $\mu \mbox{K}$ (magenta dashed line) and estimate the 
        power spectrum (blue solid circles with error bars). The estimated noise power spectrum is 
        shown slightly offset in $\ell$ to enhance clarity.}
 \label{fig:calc_validation}
 \end{figure}

We calculate both EE and BB power spectra simultaneously using the flat-sky 
approximation~\citep{flat_ps_calc}. Each simulation is run 100 times with different CMB and noise 
realization, unless otherwise noted. In this study we focus on the BB power 
spectra. The result quoted for a given $\ell$ bin is the mean of the 
100 simulations and the error bar is the standard deviation. Figure~\ref{fig:calc_validation} 
shows a validation of the process of generating CMB $Q$ and $U$ maps and estimating the underlying 
E and B-mode power spectra. No rotation due to the AHWP has been included in this validation.

For simulations that include the effects of instrumental noise we assume it is homogeneous and has a
white spectrum, and add its realization to the signal to make a combined input map. In our simulation
we use an instrumental noise per pixel of 1, 2.8, and 25 $\mu \mbox{K}$$_{\rm CMB}$ for the 150, 
250, and 410 GHz bands, respectively. Figure~\ref{fig:calc_validation} shows a validation of the noise 
generation and estimation process. 

\section{The Effect of Galactic Dust}
\label{sec:1band_sim}

If the shape of the frequency band is known then the rotation induced on the CMB signal 
alone can be calculated and compensated exactly because the spectrum of the CMB is known. 
This rotation is uniform across the sky and with the parameters given in Table~\ref{tab:sim_pars} is 
55\degr\ in the 150~GHz band. The presence of Galactic dust modifies the
intensity and angle of the net incident polarization and thus the amount of rotation induced by the
AHWP. The spectral dependence and spatial distribution of Galactic dust polarization is not precisely 
known and therefore the amount of rotation it induces can only be estimated. How big is this extra 
rotation? Can it simply be ignored because it is negligible? In the remainder of this section we assess 
these questions for the 150~GHz band.

\begin{figure}
 \begin{center}
 \scalebox{0.42}{\includegraphics[trim = 0mm 0mm 0mm 10mm, clip]{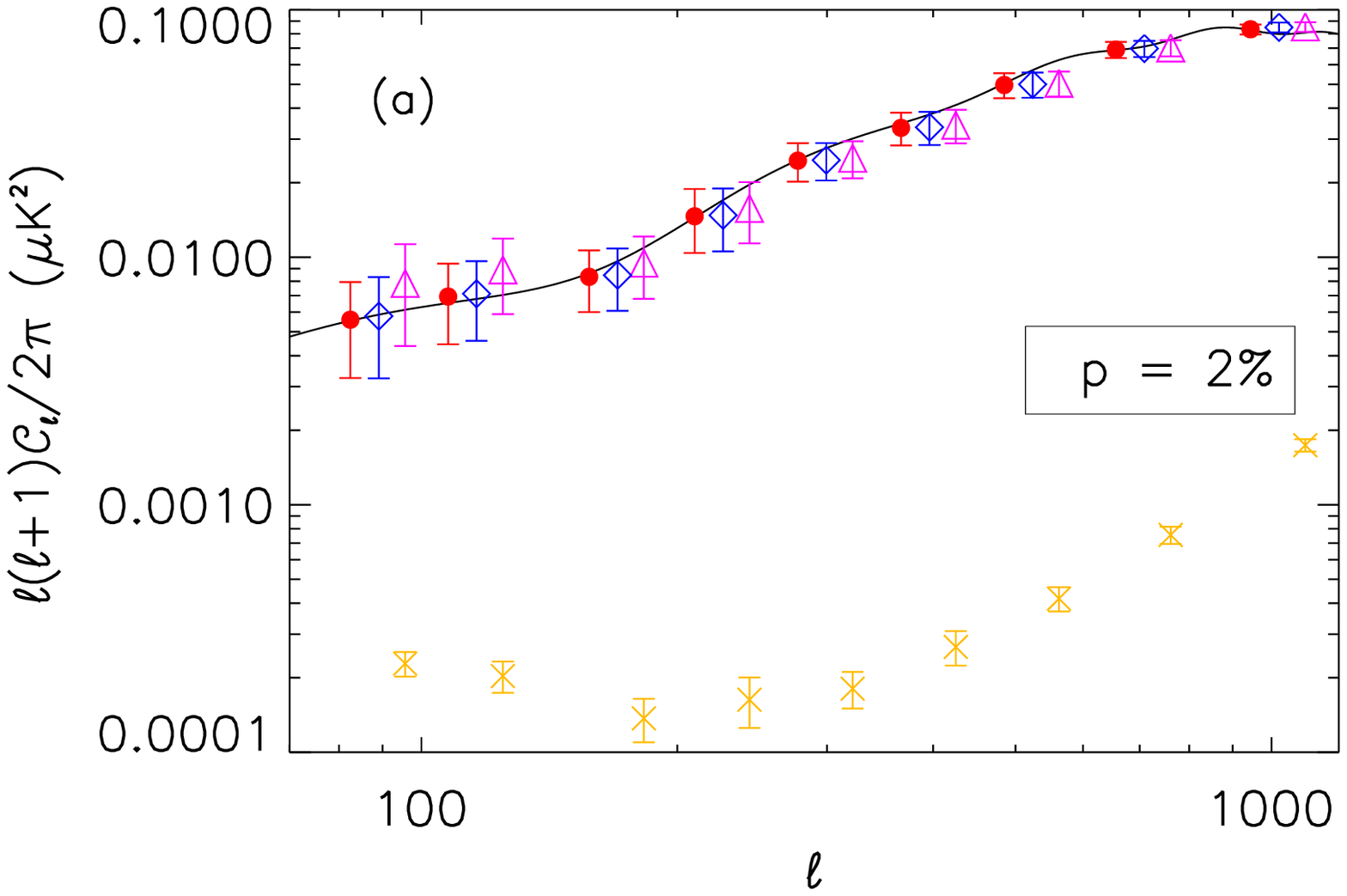}}
 \scalebox{0.42}{\includegraphics[trim = 0mm 0mm 0mm 10mm, clip]{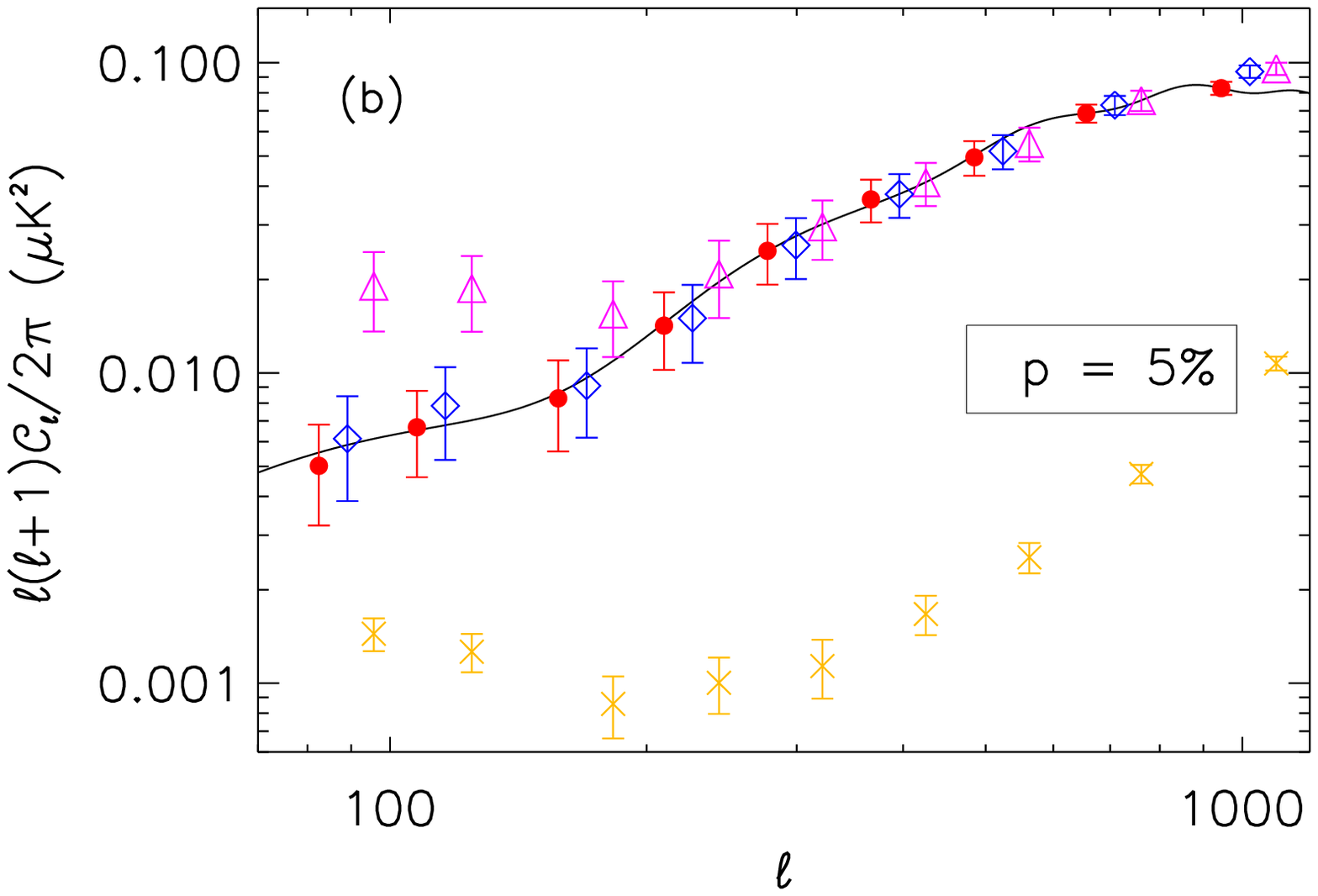}}
 \scalebox{0.42}{\includegraphics[trim = 0mm 0mm 0mm 10mm, clip]{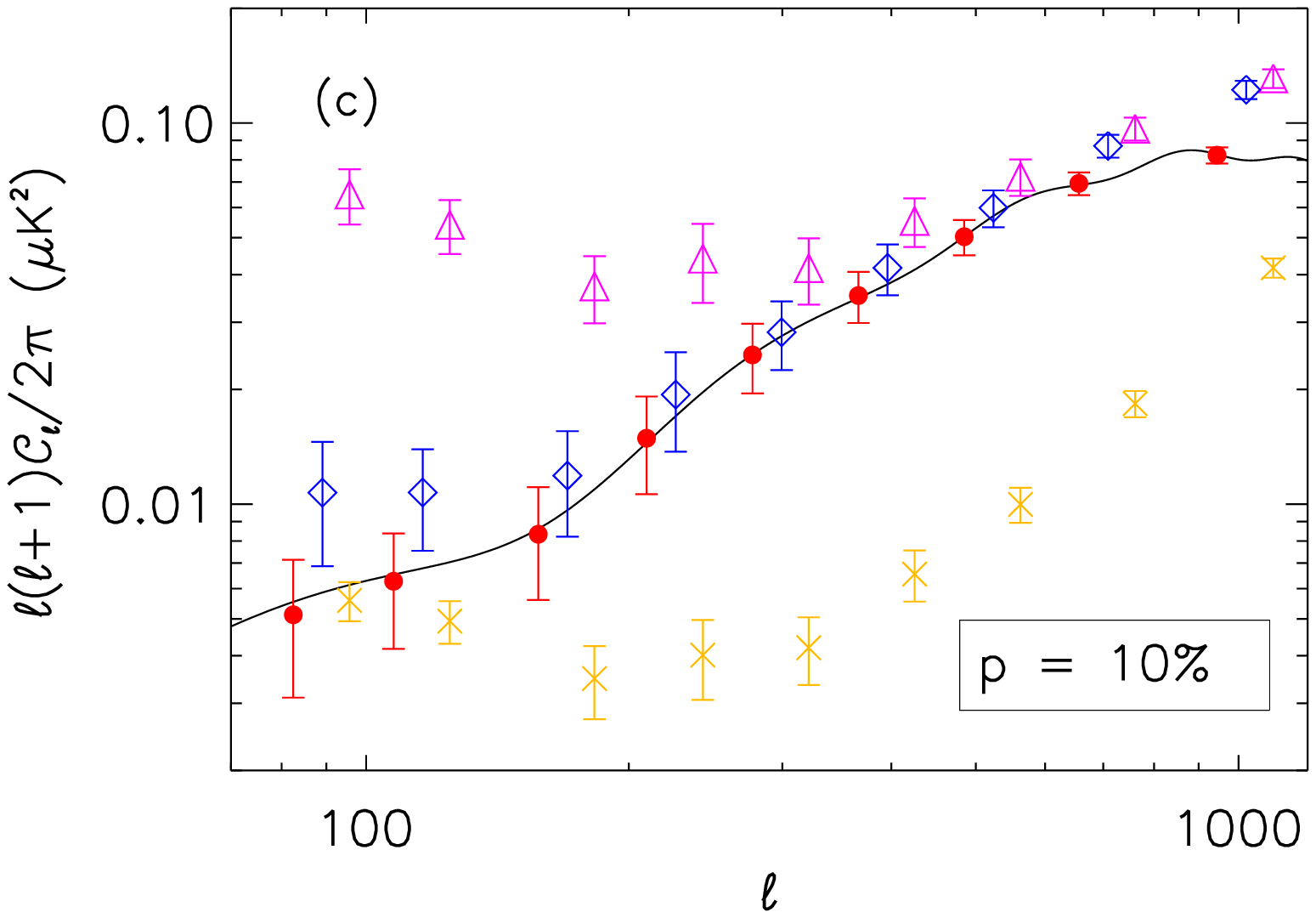}}
 \caption{Effect of rotation due to 2\%, 5\%, and 10\% polarized dust on the estimation of
  the B-mode power spectrum. The input CMB power spectrum (red solid circles) follows the underlying assumed 
  power spectrum (black solid line). The difference between the input CMB and the power spectrum of the map 
  after de-rotation by the rotation angle corresponding to the CMB alone (magenta triangles) indicates the 
  effect of polarized dust. After subtracting the dust intensity, only the {\it rotation} due to the 
  presence of dust remains (blue diamonds). The power spectrum of a map of the difference between the input
  map and the de-rotated, dust-subtracted map (yellow crosses) quantifies the effect of the rotation due to
  dust alone. Power spectra points are all calculated at the same $\ell$ bins but are shown slightly offset
  in $\ell$ to enhance clarity. To reduce clutter we only plot data for every other $\ell$ bin for $\ell > 120$.}
 \label{fig:1channel}
 \end{center}
 \end{figure}

These first simulations include CMB and dust, without instrumental noise. We calculate rotated $Q$ and $U$
maps resulting from passing the total (CMB+dust) incoming polarization through the AHWP.  We then 
`de-rotate' the maps by the calculated rotation angle for CMB only, simulating ignorance of the effects 
of the dust foreground on the rotation. We subtract the input dust $Q$ and $U$ maps
from the de-rotated map to acknowledge the presence of the effects of dust {\it polarized intensity} on the 
total $Q$ and $U$ maps. Note however that the effect of {\it rotation due to dust polarization}, 
which is a consequence of the AHWP, is left in the map.  We then calculate the angular power spectrum of 
the resulting maps for 2\%, 5\%, and 10\% of dust polarization (see Figure~\ref{fig:1channel}). 
For the fiducial value of $r=0.05$ ignoring the effect of rotation introduces noticeable bias in the 
estimation of the CMB power spectrum for 10\% dust polarization but not for 2\% dust polarization. For
5\% dust polarization the bias is only noticeable in the lowest two and the highest $\ell$ bins.

\section{Removing AHWP Induced Rotation in Dust Subtraction}
\label{sec:multi_band_pipeline}

In the previous section we showed that for levels of polarized dust of more than 5\% the effect of 
the rotation due to dust in the AHWP cannot be ignored. In this section we employ a simple
form of dust subtraction in an attempt to correct for the rotation. In this approach we 
make two initial assumptions in order to extract the dust frequency scaling information: 
\begin{enumerate}
\item the signal at the 150 GHz band is dominated by the CMB and dust can be neglected; 
\item the signal at the 410 GHz band comes entirely from dust and CMB can be neglected;
\end{enumerate}

We prepare the total (CMB+dust) rotated polarization maps in 150, 250, and 410~GHz bands and add
noise in the map domain. We then calculate the polarization intensity maps and the signal RMS in all
three bands assuming that the noise RMS is exactly known. Following assumption 1 the map RMS of 
CMB at 150 GHz is known. We extrapolate the CMB level to 250~GHz and calculate the map RMS for dust 
at 250~GHz. Following assumption 2 we also obtain the map RMS for dust at 410~GHz. Using the dust 
levels at 250 and 410~GHz we fit a gray body dust model, given by a power law multiplied by an 18 K 
blackbody. The power law spectral index is taken to be uniform across the entire simulated sky area. 
The top hat spectral response of the instrument is assumed to be precisely known (we address 
uncertainties in the bandpass in Section~\ref{sec:band_sims}).
The fitted dust model is used to calculate the level of dust at the 150~GHz band, extrapolated from the 
410 GHz map, and to calculate and correct for the rotation angle at this band due to the combination of 
dust and CMB. We make a final map that contains an estimate of the CMB alone after corrections for both 
dust polarized intensity and rotation induced by the AHWP. We calculate the power spectrum of this map 
including subtraction of an estimate of the noise spectrum. For an estimate of the noise spectrum we use 
the known input RMS. As a test of the entire pipeline we run it with no dust and no noise and validate 
that the extracted power spectrum agrees with the input CMB power spectrum. 
 
\begin{figure}
 \begin{center}
 \scalebox{0.42}{\includegraphics[trim = 0mm 0mm 0mm 10mm, clip]{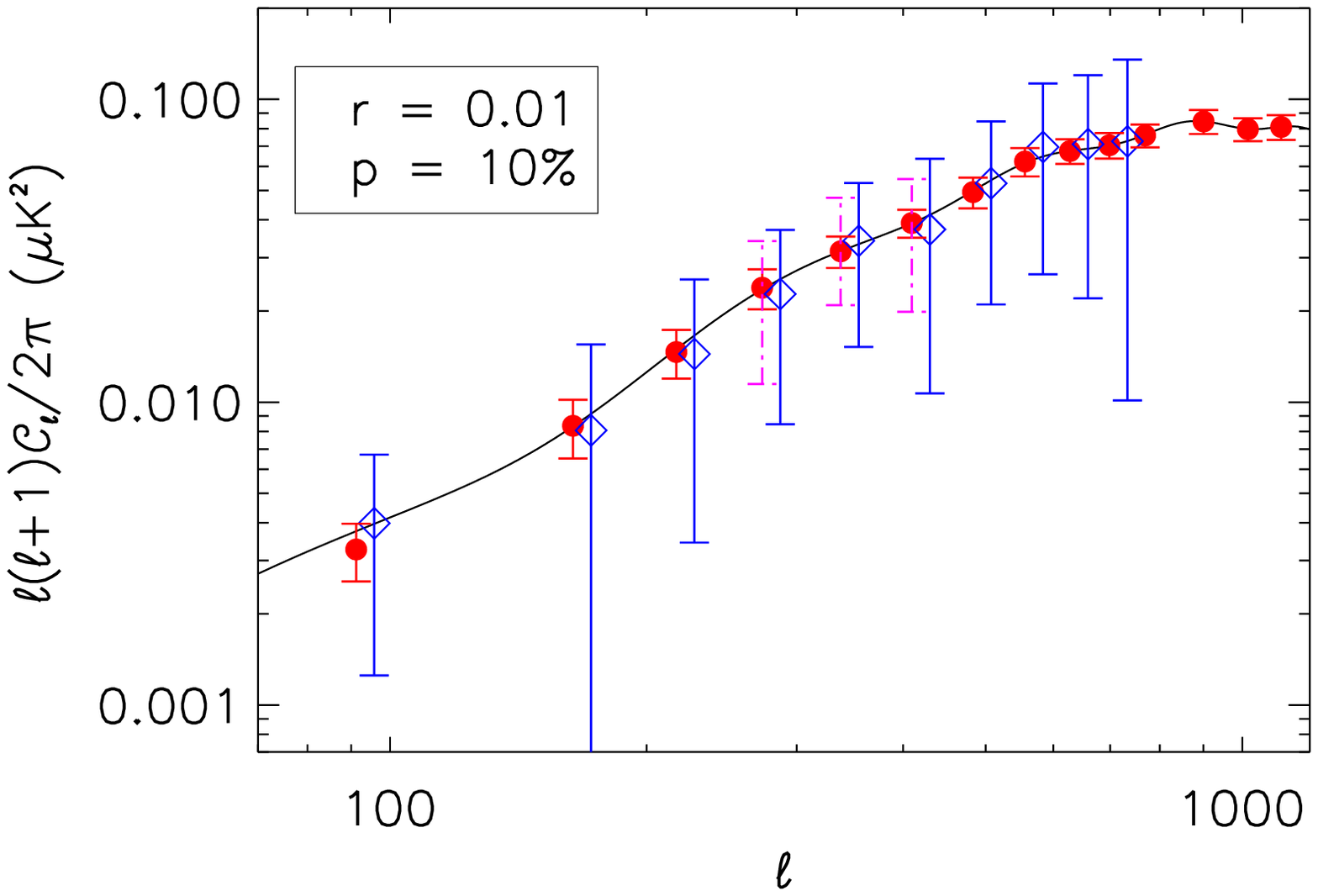}}
 \scalebox{0.42}{\includegraphics[trim = 0mm 0mm 0mm 10mm, clip]{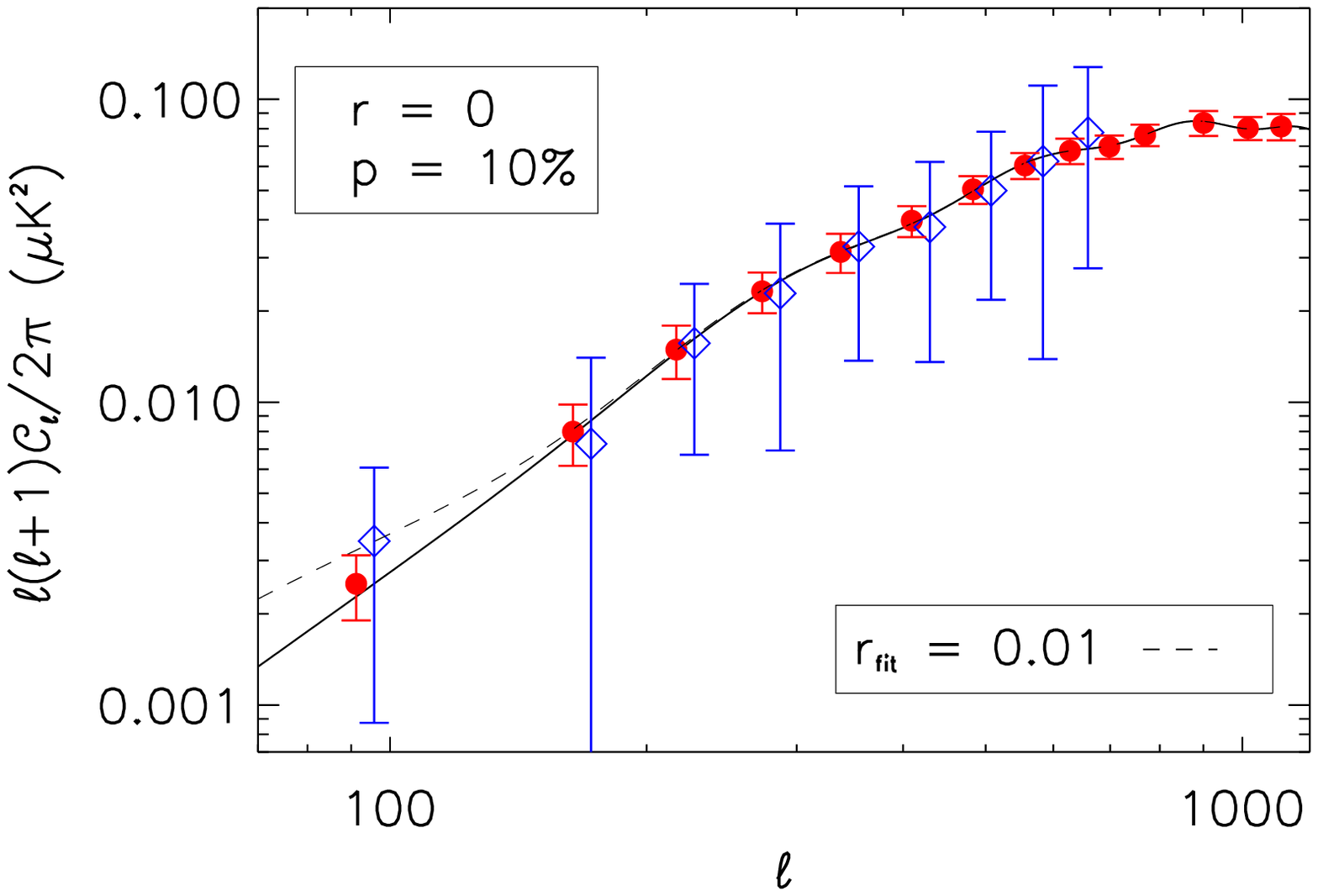}}
 \caption{Comparison between the underlying CMB model (black line), the input CMB (red circles, 
 error bars include only cosmic variance) and the estimate of the CMB power spectrum using 
 a map in which the effects of both dust polarized intensity and dust induced extra rotation has been 
 accounted for (blue diamonds). The final error bars are on average 30\% larger than those
 due to cosmic variance \textit{and} instrument noise (magenta dot-dashed error bars, only plotted in three mid-$\ell$ 
 range bins for clarity). With 10\% dust polarization, a crude dust subtraction algorithm (see text) can 
 account for the rotation induced in the AHWP with an $r$ value as low as 0.01 (top panel). In the 
 bottom panel the input $r$ is zero and the underlying CMB spectrum has only a lensing signal; Galactic dust 
 is 10\% polarized. The estimated CMB spectrum has a best fit $r=0.01$. To reduce clutter we only 
 plot data at every other $\ell$ bin for $\ell > 800$.}
 \label{fig:pipeline_limit}
 \end{center}
 \end{figure}

For each set of 100 simulations, we determine whether the final estimated CMB power spectrum is biased
or not. The power spectrum is conservatively assumed biased if the mean power estimated in {\it any}
$\ell$ bin is outside of the 1-$\sigma$ cosmic variance error bar. We find that for the nominal noise 
levels, $r=0.01$ or above, and all dust polarized fractions at or below 10\% the dust subtraction 
procedure recovers an unbiased estimate of the B-mode power spectrum. The results for 10\% dust 
polarization fraction are shown in Figure~\ref{fig:pipeline_limit}. Only data points with signal to 
noise ratio (SNR) $> 1$ are plotted. For $r=0.009$ and lower,
and 10\% dust polarization fraction, we find that the recovered B-mode power spectrum is 
biased at the lowest $\ell$ bin. When the dust polarization is lower the $r$ level that can be recovered
without bias is higher because the higher relative noise at 410 GHz has a larger effect on the CMB estimate
at 150 GHz. For 2\% dust polarization fraction, we can recover $r$ as low as 0.02.

We also use a different approach to 
quantify the bias caused by the dust subtraction procedure. We run 100 simulations with an input of $r=0$ 
and 10\% dust polarization fraction. We fit a non-zero $r$ to the difference between the 
estimated and input CMB power spectrum at the 
lowest $\ell$ bin, while keeping the shape of the primordial B mode 
signal. We consider this fit as the lower limit of $r$ value we can detect using this dust subtraction 
method. We find a best fit with $r=0.01$, which is close to the result we found earlier.

The error bars shown in Figure~\ref{fig:pipeline_limit} include only cosmic variance for the input CMB, but 
include the total error on the recovered CMB power spectrum (including the effect of correcting for 
the rotation). On average the total error is larger by 30\% compared to the error from the combination 
of cosmic variance and instrument noise (Fig.~\ref{fig:pipeline_limit}, upper panel).

\section{Uncertainty in Detection Band and High Frequency Spectral Response}
\label{sec:band_sims}

So far we assumed that the spectral response of the instrument is known. Only the frequency scaling 
of Galactic dust is determined from the fit. Uncertainty in the spectral response leads to 
uncertainty in the amount of rotation induced by the AHWP. To assess the level of this effect
quantitatively we assume a top-hat band shape that is characterized by two parameters, center 
and width. We simulate CMB and dust signals with {\it nominal} bands, and analyze the maps using the 
dust subtraction algorithm discussed in the previous section but assuming varying band-widths, or varying 
band-centers (but not both simultaneously). No instrumental noise is included. All simulations have 
10\% polarized dust and $r=0.05$. We search for the level of shift in band-center or change in width that 
leads to bias in the estimation of the final CMB spectrum. We use the same criterion for bias as described 
in Section \ref{sec:multi_band_pipeline}.

\subsection{Shift of Band-Center}

\begin{figure}
\begin{center}
\scalebox{0.42}{\includegraphics{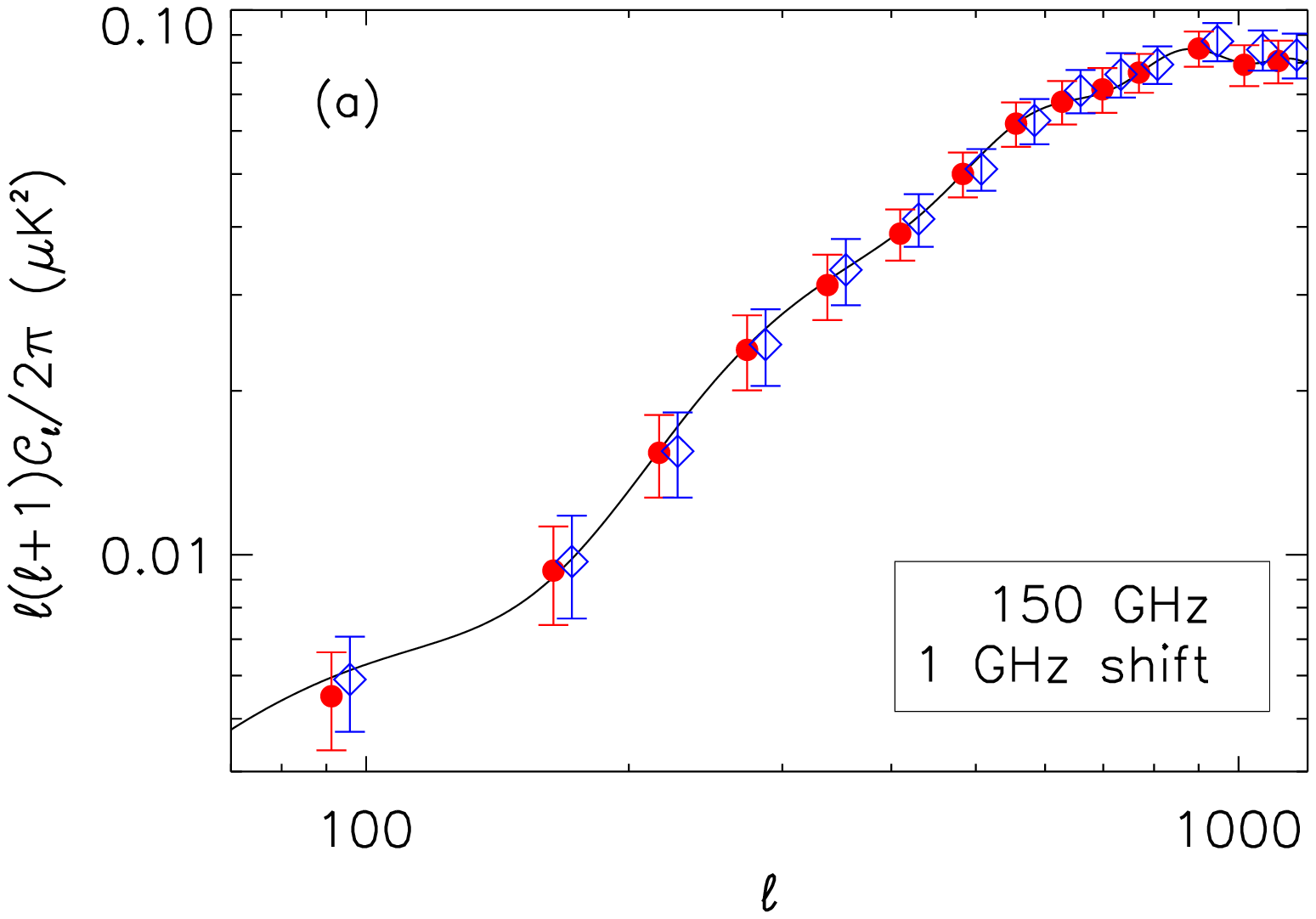}}
\scalebox{0.42}{\includegraphics{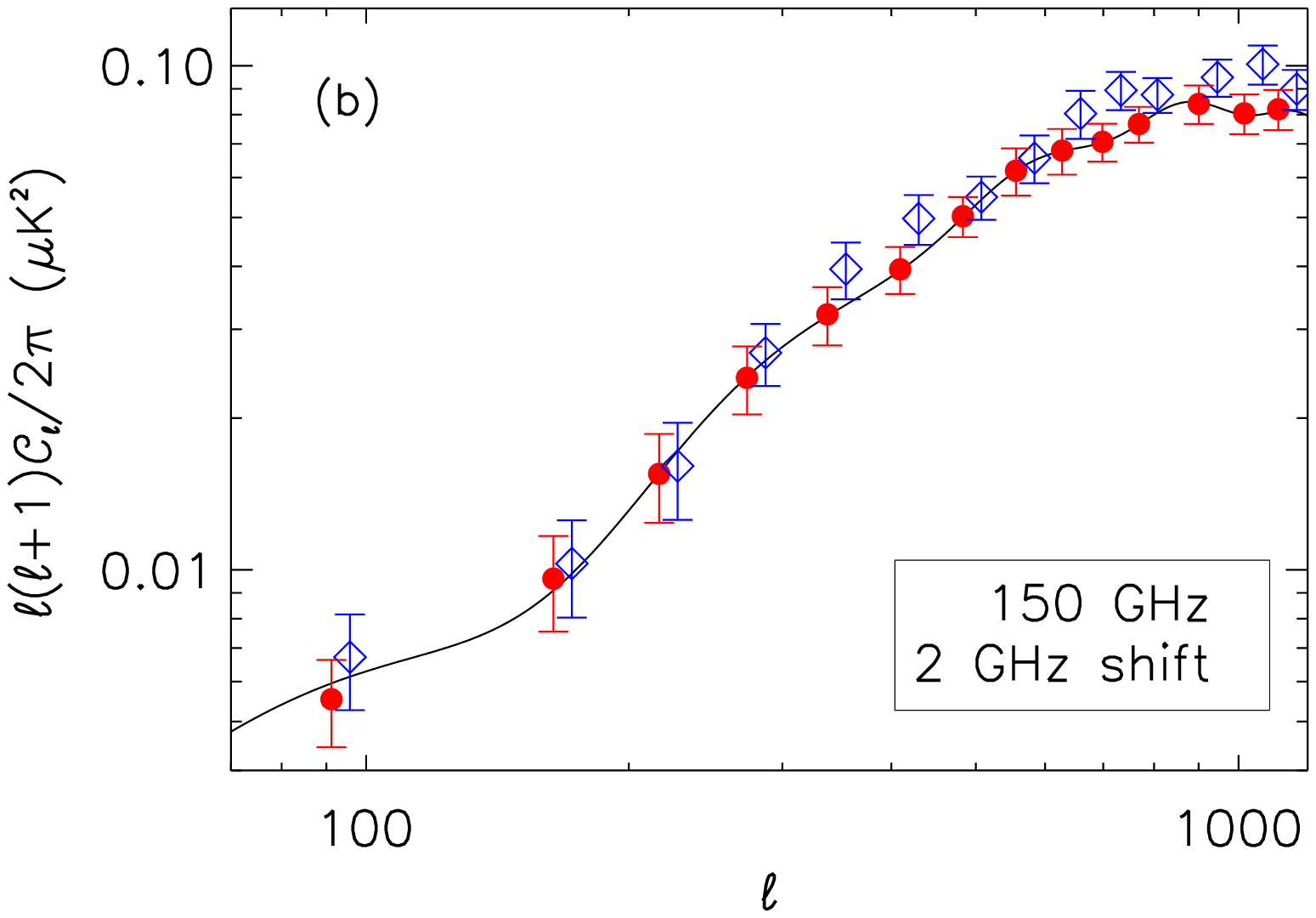}}
\scalebox{0.42}{\includegraphics{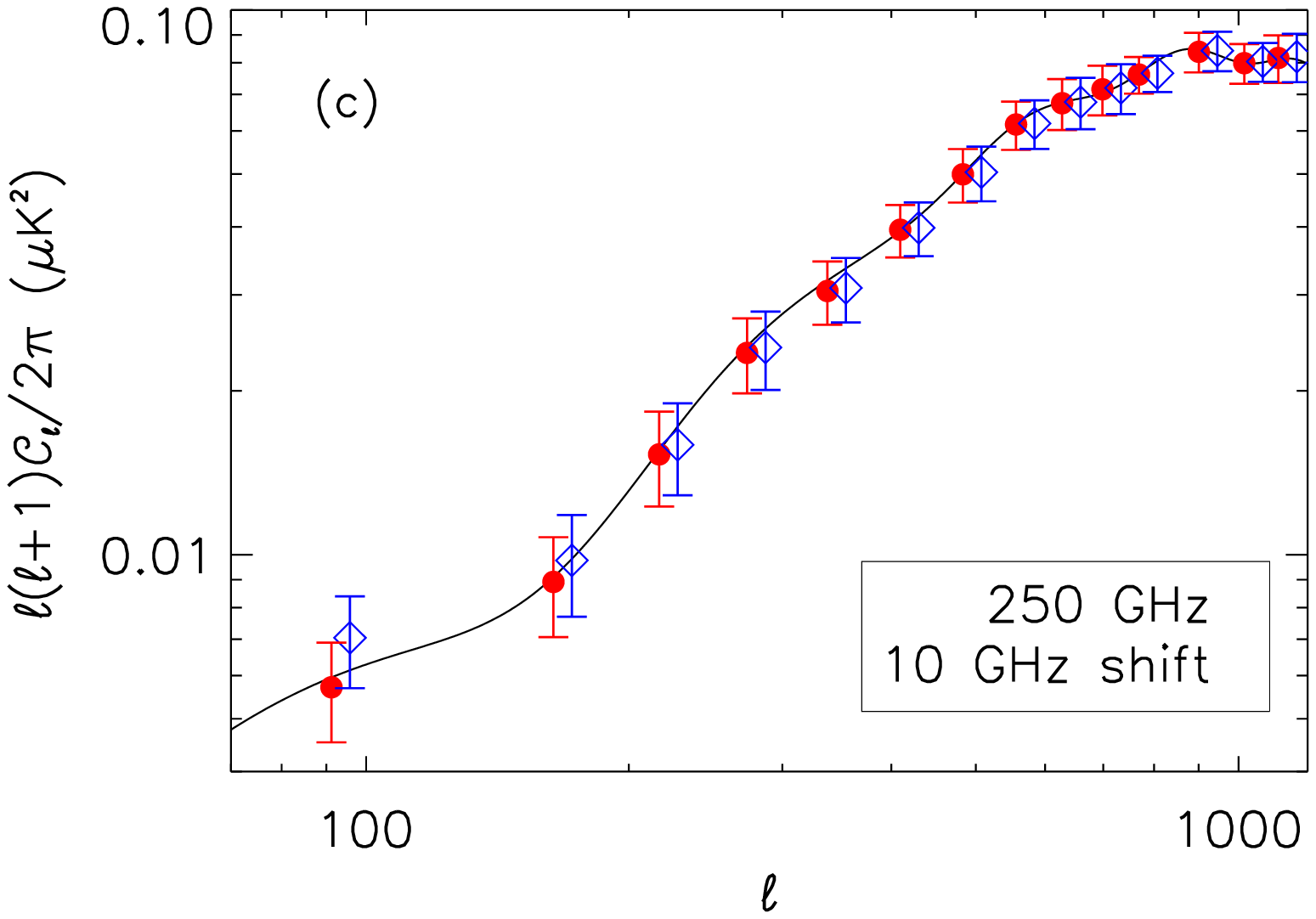}}
\scalebox{0.42}{\includegraphics{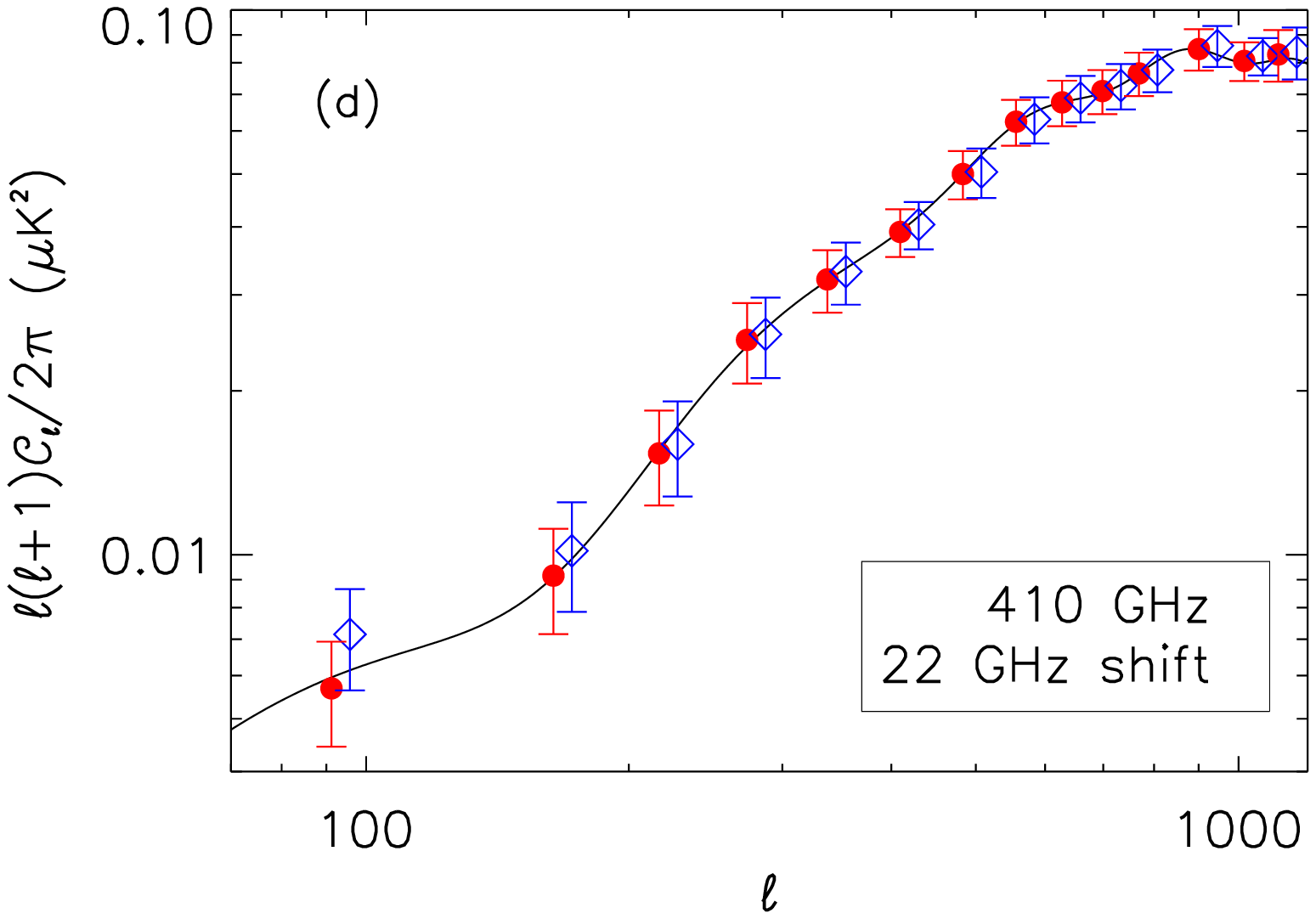}}
\caption{Effect of band-center shift of 1~GHz of the 150 GHz band (panel a), 2~GHz of the 150~GHz band 
(panel b), 10~GHz of the 250 GHz band (panel c) and 22~GHz of the 410~GHz band (panel d), respectively. 
In panel (b) excess power at high $\ell$ comes from mixing of E and B-mode. In panels (c) and (d) it 
comes from misestimate of the effects of dust. Figure
symbols are the same as in Figure \ref{fig:pipeline_limit}. }
\label{fig:band_center}
\end{center}
\end{figure}

Simulations are carried out by shifting only one band-center, keeping the other two fixed at their nominal
values. We find that 
shifts of more than 1, 9, and 20 GHz for the 150, 250, and 410 GHz bands, respectively, lead to biased
power spectra. The limit for shift of the 150 GHz band is due to mixing between 
E and B modes: an error in band-center leaves the CMB slightly rotated after correction for the AHWP is 
applied (using the nominal band) and thus a portion of the E-mode signal is mixed into the B-mode 
signal. This is apparent in 
Figure~\ref{fig:band_center} (panel b), which shows a 2~GHz shift for the 150 GHz; the bias is primarily 
at high $\ell$. The limits on the 250 and 410~GHz bands mainly come from misestimate of rotation due to 
dust, but because dust is not dominant at the 150~GHz band the requirement is less stringent. Panels (c) 
and (d) in Figure~\ref{fig:band_center} show shifts of 10 and 22~GHz for the 250 and 410~GHz bands, in 
which bias due to dust is found only at the lowest $\ell$ bin.

\subsection{Misestimate of Band-Width}
\label{sec:band_width}

\begin{figure}
\plotone{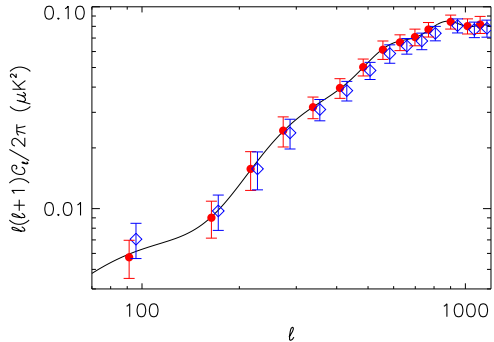}
\caption{A 1~GHz increase in the assumed band-width of the 150~GHz band relative to the nominal width
leads to an overestimate of the level of dust in this band and thus to a misestimate of the rotation 
due to dust. This leads to excess power at low $\ell$. The misestimate of the band-width also results
in an underestimate of the CMB power which leads to a small (less than 1-$\sigma$) power deficit at 
high $\ell$ bins. Figure symbols are the same as in Figure~\ref{fig:pipeline_limit}.}
 \label{fig:band_width}
 \end{figure}

Simulations are carried out by changing one band-width at a time, keeping the other two fixed at their 
nominal values. We find that a misestimate of band-width by more than 
0.8, 2, or 14~GHz for the 150, 250, and 410 GHz band, respectively, exceeds the 
criterion for no bias. The result for the 150~GHz band with a change of 1~GHz in width
is shown in Figure \ref{fig:band_width}. The cause of bias in any of the bands is a misestimate 
of total power detected at the particular band thus a misestimate of both the polarized signal
intensity and the rotation due to dust. For this reason the bias is largest at the lowest $\ell$ bins. 

\subsection{Effects of High Frequency Spectral Leak}

\begin{figure}
\plotone{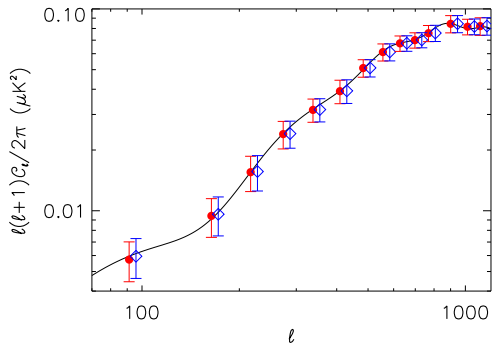}
\caption{The effect of a broad high frequency spectral leak (see text for details) is negligible with 
$r=0.05$ and 10\% dust polarization. Figure symbols are the same as in Figure~\ref{fig:pipeline_limit}.}
\label{fig:high_f_leak}
\end{figure}

Dust intensity is rising  up to $\sim2$~THz. A higher than expected and unknown instrumental response 
at out-of-band frequencies, which is called a `spectral leak', may bias the subtraction of the dust 
signal and by extension the estimate of the underlying CMB signal. We simulate two specific leaks, 
which are both top-hat in shape, a narrow leak between 1750 and 1850~GHz and a broad leak between 500 
and 2000~GHz. For both cases the power in the leak is adjusted to be 0.1\%, 1\% and 1\% of the 
in-band power for the 150, 250, and 410~GHz bands, respectively, as measured with a 300 K blackbody 
source. These values are readily achievable experimentally~\citep{polsgrove_thesis}.
We properly include the change in the refraction indices of sapphire with frequency~\citep{Loewenstein73}. 
Maps are prepared with signals that include power in the leak, but are analyzed, including the steps 
of dust subtraction, assuming no knowledge of the leak. Instrumental noise is not included in  the 
simulation. For both cases we find no biases in the estimate of the final CMB power spectrum.  
Figure \ref{fig:high_f_leak} shows the case for the broad leak.

\section{Discussion and Summary}

The spectral response of an achromatic half-wave plate may induce bias in 
the estimation of polarized signals. We analyze the level of such bias in the context of 
measurements of the B-mode signal of the CMB in the presence of Galactic dust, the dominant source of
foreground emission in cases of interest here. For concreteness, we use the specific experimental configuration 
corresponding to the EBEX balloon-borne experiment. 
 
For the area of sky considered we find that with reasonable assumptions about the magnitude and spectral 
shape of dust, the effects of rotation induced by the AHWP are only appreciable when dust is polarized at 
a level of about 5\% and above and the tensor-to-scalar ratio $r$ is less than $\sim0.05$. In the regime
when the effects of rotation are appreciable, even a crude process of dust estimation and 
subtraction mitigates the effects of AHWP rotation to below detectable levels. For 
example, using the crude dust subtraction process
we find no bias in the estimation of the B-mode power spectrum for dust polarization fraction 
as large as 10\% and $r$ as low as $0.01$. For 2\% dust polarization fraction, $r$ of 0.02 or higher is
recovered without bias. We also find that the dust subtraction causes the power spectrum error 
bars to increase by a modest 30\% on average.

Employing the same dust estimation and subtraction process, but now assuming 
errors in knowledge of the experiment's detection band-center and band-width, we find the 
accuracy with which these need to be measured. For example, for the particular experimental 
configuration considered, we found that band-center and band-width of the 150 GHz band need
to be determined to better than 1 and 0.8~GHz, respectively. It is possible that this requirement 
may not need to be as stringent if a more sophisticated foreground estimation and subtraction process
is used. This research is ongoing. 

We explore the sensitivity of the particular experimental configuration to high frequency 
spectral leaks. Using a rejection level that is readily achievable experimentally we show
that spectral leaks are not expected to pose challenges for the operation with an AHWP. 

The analysis and subtraction approach discussed in this paper are applicable to other 
optical elements for which polarization rotation is a function of frequency. For example, 
~\citet{OBrient_thesis} describes a broadband, mm-wave detection technique that is based on 
sinuous antenna. It is well documented that such antennas change the phase response 
of polarized signals, and that this effect is frequency dependent. Thus they exhibit 
fundamentally the same behavior as an AHWP. Our methods and approach apply to 
such cases. 

\acknowledgments

This research project is supported by a National Science Foundation (NSF) grant 00011640. We are
 greatly thankful for the computing resources provided by the Minnesota Supercomputing
 Institute. EBEX is supported through NASA grants NNX08AG40G and NNX07AP36H. C.Baccigalupi and 
 S.Leach acknowledge travel support from the PD51 INFN grant. A.Jaffe acknowledges the support 
 from the STFC in the UK.

\bibliography{hwp_paper}
\bibliographystyle{apj}
\end{document}